\documentclass[pra,twocolumn,showpacs,floatfix]{revtex4}
\usepackage{graphicx}
\begin{document}

\title{Persistent currents in a Bose-Einstein condensate
in the presence of disorder}
\author{M. \"Ogren and G. M. Kavoulakis}
\affiliation{Mathematical Physics, Lund Institute of
Technology, P.O. Box 118, SE-22100 Lund, Sweden}
\date{\today}

\begin{abstract}

We examine bosonic atoms that are confined in a toroidal,
quasi-one-dimensional trap, subjected to a random potential.
The resulting inhomogeneous atomic density is smoothened for
sufficiently strong, repulsive interatomic interactions.
Statistical analysis of our simulations show that the gas
supports persistent currents, which become more fragile due to
the disorder.

\end{abstract}
 \maketitle

\section{Introduction}

Superfluidity is one of the most fascinating problems in
physics. Superfluidity includes a whole class of phenomena that
appear in various quantum systems \cite{Leggettsf}, such as
liquid He, superconductors, and nuclei. More recently, a lot of
effort has been put on the study of superfluid properties of
dilute vapors of trapped atoms, including, for example, vortex
states \cite{Dal1,Dal2,Kett,Corn}, transport properties
\cite{Kettsf}, etc. Although these gases are very dilute, they
are superfluid because of their extremely low temperature.
Their diluteness makes cold gases of atoms an ideal system for
testing superfluid properties. For example, their theoretical
description is much easier as compared to other superfluids,
like e.g., liquid Helium. Furthermore, they can be manipulated
in numerous ways, including the form of the trapping potential,
and their coupling constant.

In the present study we examine the stability of persistent
currents in trapped superfluid atoms that are confined in
toroidal, quasi-one-dimensional traps. While conservation laws
imply trivially the conservation of momentum/angular momentum
even in a moving/rotating classical system, the crucial
question is the robustness of the current-carrying state(s)
against impurities or anisotropies; it is exactly this feature
that distinguishes a quantum from a classical system.

Many theoretical studies have examined this problem, see, e.g.,
Refs.\,\cite{pl,mf,Huang,Giorgini1,Giorgini2,Tsubota,Graham,LSP}.
Experimentally, numerous recent studies have considered this
question in connection with trapped atomic gases
\cite{Wiersma1,Wiersma2,Bouyer1,Bouyer2}. In another class of
experiments -- which are closely connected with the present
problem -- the propagation of Bose-Einstein condensed atoms in
magnetic waveguides has also been examined thoroughly
\cite{Mtr}.

Interestingly, the problem that we study here is not only a toy
model that provides answers on some non trivial questions, but
also it is directly applicable to experiments that have created
Bose-Einstein condensates of atoms in toroidal traps
\cite{Kurn}. Furthermore, the stability of supercurrents in
these mesoscopic systems may have revolutionary technological
applications \cite{appl}; similar arguments may also be
applicable to other (nanoscale/mesoscopic) systems, such as
quantum rings and quantum dots \cite{RM}.

In what follows we first describe our model in Sec.\,II. In
Sec.\,III we consider the effect of a random potential on the
atomic density of a static cloud. In Sec.\,IV we discuss how
one may understand the existence of persistent currents as a
result of the metastability of the current-carrying state.
Section V presents the results for the effect of disorder on
the stability of superflow within the mean-field approximation.
In this section, we consider many random potentials and analyze
statistically our results. In Sec.\,VI we solve the same
problem within the Bogoliubov approximation. Finally, in
Sec.\,VII we discuss our results and give some general
conclusions.

\section{Model}

In our model we consider a tight toroidal trap, which freezes
the transverse degrees of freedom of the motion of the atoms.
Further, we assume for the atom-atom collisions the usual
contact potential, $V_{\rm int}({\bf r}-{\bf r}') = U_0
\delta({\bf r}-{\bf r}')$ with $U_0 = 4 \pi \hbar^2 a_{\rm sc}
/M$. Here, $a_{\rm sc}$ is the scattering length for elastic
atom-atom collisions and $M$ is the atomic mass. Since the
transverse degrees of freedom are frozen, the three-dimensional
field operator may be written as $\sqrt{N/(RS)} \, \hat
\Psi(\theta)$, where $\hat \Psi(\theta)$ is the field operator
associated with the motion of the atoms along the torus. Here
$\theta$ is the azimuthal angle, $N \gg 1$ is the atom number,
$R$ is the radius of the torus, and $S$ is the cross section of
the torus (with $R \gg \sqrt S$). Therefore, the Hamiltonian of
our quasi-one-dimensional system is \cite{Ueda,GMK}
\begin{eqnarray}
  \hat H / N =  \int {\hat \Psi}^{\dagger}(\theta)
  \left[ - \frac {\hbar^2} {2MR^2}
  \frac {\partial^2} {\partial \theta^2} + V(\theta)
  \right]
  {\hat \Psi}(\theta) \, d \theta +
  \nonumber \\
+ \frac 1 2 2 \pi n_0 U_0 \int
  {\hat \Psi}^{\dagger}(\theta) {\hat \Psi}^{\dagger}(\theta)
  {\hat \Psi}(\theta) {\hat \Psi}(\theta) \, d \theta,
\label{Ham1}
\end{eqnarray}
where $V(\theta)$ is the external potential, which is taken to
be piecewise constant and random (see bottom graph in Fig.\,1).
Finally, $n_0$ is the average atom density, $n_0 = N /(2 \pi R
S)$.

Within the mean-field approximation the condensate order
parameter $\Psi$ satisfies the nonlinear equation
\begin{eqnarray}
  - \frac {\partial^2 \Psi} {\partial \theta^2}
  + V(\theta) \Psi(\theta)
  + 2 \pi \gamma|\Psi(\theta)|^2 \Psi
  = \mu \Psi,
\label{gpee}
\end{eqnarray}
where we have set $\hbar = 2M = R = 1$; $V(\theta)$ and the
chemical potential $\mu$ are measured in units of the kinetic
energy $T = \hbar^2/(2 M R^2)$. The ratio between the
interaction energy and the kinetic energy is equal to $\gamma =
n_0 U_0 / T$. This parameter is also equal to $\gamma = 4 N
a_{\rm sc} R/S$.

The mean-field approximation is valid when the dimensionless
quantity $\gamma$ is $\ll N^2$ \cite{KYOR}. Further, the limit
of weak interactions is achieved when $\gamma \ll 1$, while the
Thomas-Fermi limit is achieved when $1 \ll \gamma \ll N^2$. For
the typical values of $\gamma$ that we consider in the present
study, the mean-field approximation is applicable, and the
healing length $\xi$ is on the order of the radius of the
torus, since $\xi/R = \gamma^{-1/2}$. The opposite
Tonks-Girardeau limit of impenetrable bosons is achieved when
$\gamma$ becomes of order $N^2$ \cite{KYOR}, where $\xi/R \sim
1/N$.

\section{Density profile in the presence of a random potential}

As a first step, we consider the static profile of the gas in
the presence of the random potential $V(\theta)$. We choose the
length scale $d$ of variation of $V(\theta)$ to be $R/10$, as
in the experiment of Ref.\,\cite{Wiersma1}. In this experiment,
the axial size of the condensate was $\approx 110$ $\mu$m, the
radial size was $\approx 11$ $\mu$m, the smallest length scale
of each ``speckle" was $\approx 10$ $\mu$m, and the average
distance between speckles was $\approx 20$ $\mu$m.

There are four energy scales in our problem, namely the kinetic
energy associated with the torus $\hbar^2/(2MR^2)$, the
characteristic depth of each separate potential ${\tilde V}_0$, the
zero-point kinetic energy $\hbar^2/(2 M d^2)$, and the typical
interaction energy $n_0 U_0$. Since we choose $d = R/10$,
therefore $\hbar^2/(2 M d^2)$ is two orders of magnitude larger
than $\hbar^2/(2 M R^2)$. We also choose ${\tilde V}_0$ to be of the
same order as $\hbar^2/(2 M R^2)$, which implies that the
lowest state, as well as the low-lying excited (eigen)states of
the non-interacting problem are localized around the minima of
the random potential, but there is also a significant overlap
between the maxima in the density, as shown in Fig.\,1 for
$\gamma=0$. If one chooses the zero of the energy to be at the
maximum of the random potential, quantum mechanics implies that
in the effectively one-dimensional problem that we consider
here there is always at least one bound state, and thus at
least one exponentially localized/delocalized state
\cite{Jaffe,JK}.

Under these conditions, depending on $\gamma$, there are three
different regimes. For zero/weak interactions, $\gamma \ll 1$,
the order parameter is equal/close to the lowest eigenstate of
the external potential. As $\gamma$ increases, the density
variations get suppressed. When $\gamma \gg 1$, the density
becomes homogeneous, as shown in Fig.\,1.

\begin{figure}[t]
\includegraphics[width=7.2cm,height=4.5cm]{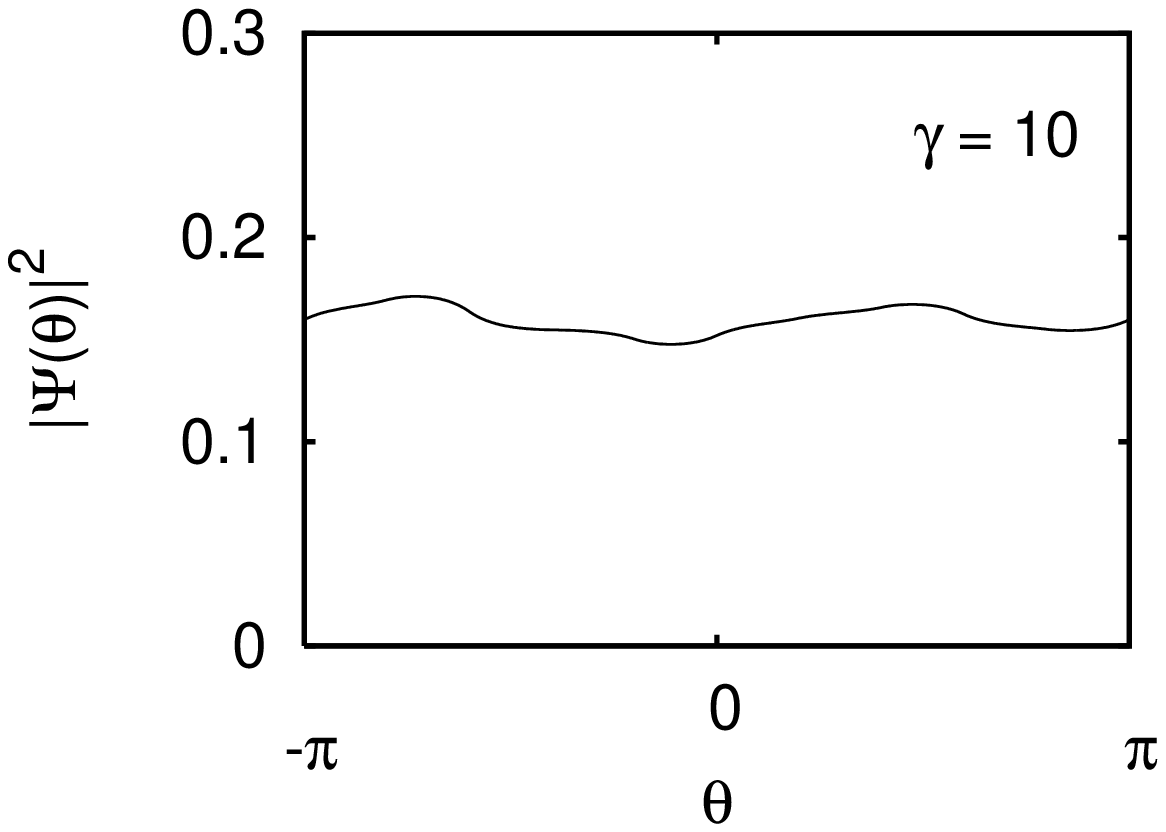}
\includegraphics[width=7.2cm,height=4.5cm]{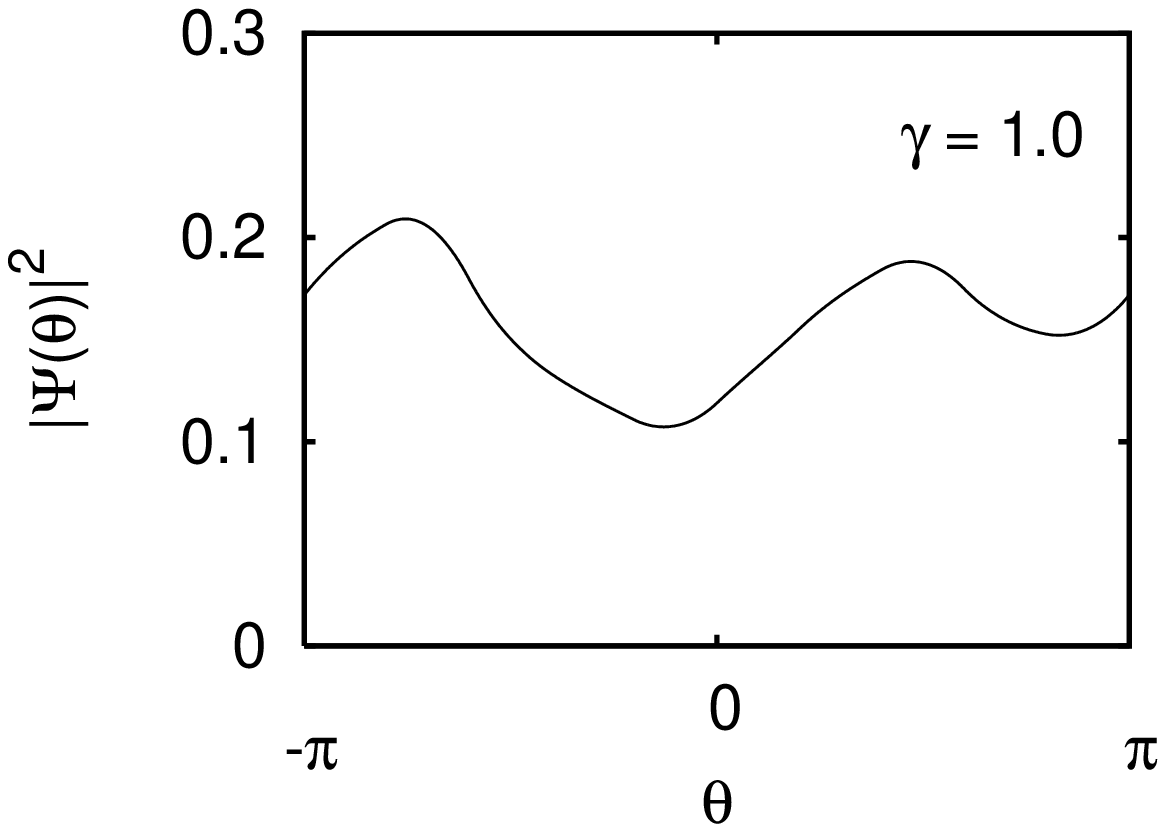}
\includegraphics[width=7.2cm,height=4.5cm]{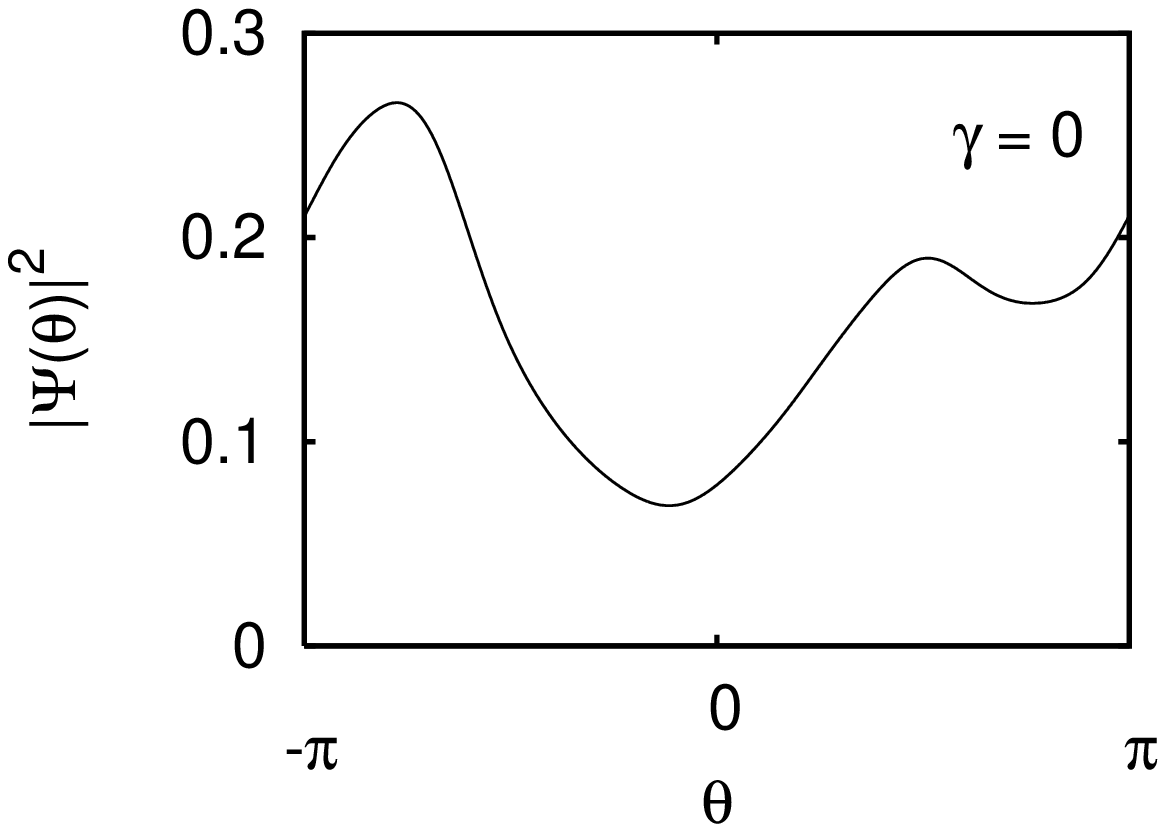}
\includegraphics[width=7.2cm,height=4.5cm]{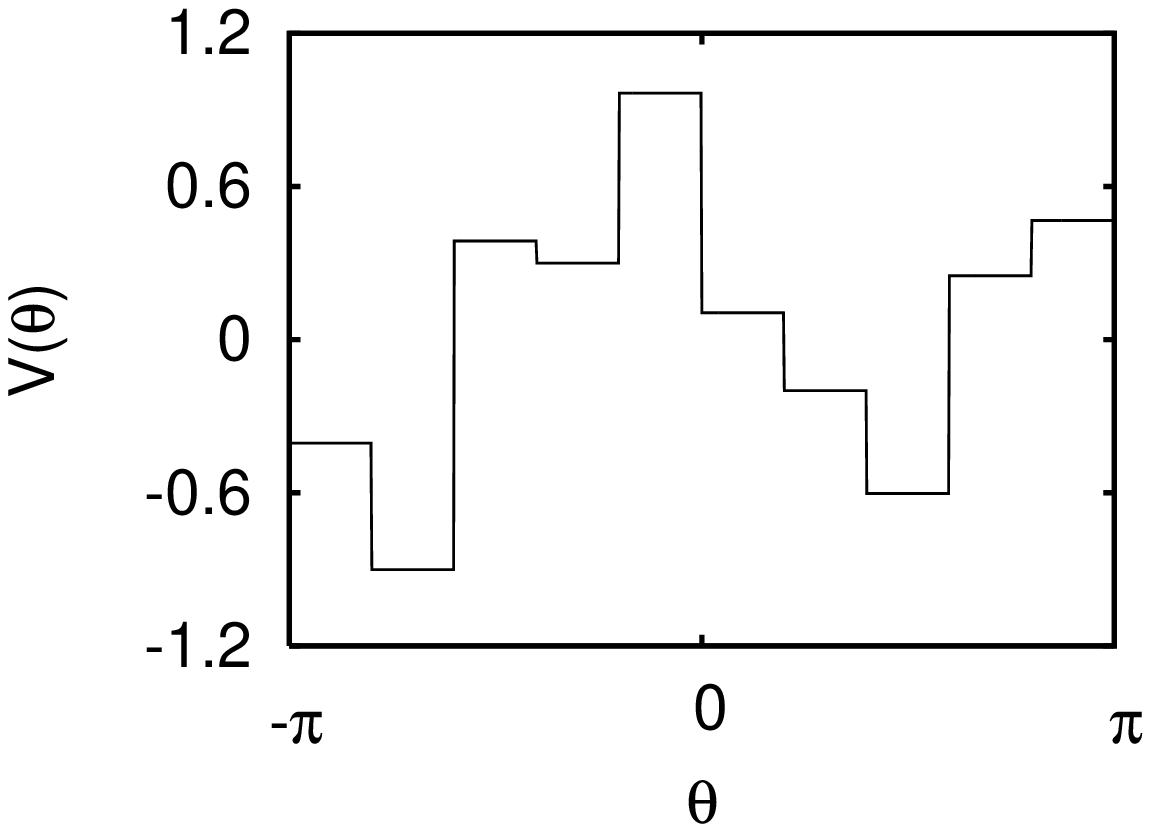}
\caption[]{The density $|\Psi(\theta)|^2$ of Eq.\,(\ref{gpee})
for three values of the coupling constant $\gamma = 0.0, 1.0$,
and 10.0, for the random potential $V(\theta)$ shown in the
bottom graph.}
\label{FIG1}
\end{figure}

\section{Stability of persistent currents}

We turn now to the stability of persistent currents. Before we
consider any disorder, it is instructive to see how one
understands the existence of persistence currents for a
constant potential $V(\theta)$.

The basic idea is that for sufficiently strong interactions,
the dispersion relation ${\cal E}(l)$, i.e., the energy per
atom as function of the angular momentum per atom, develops a
barrier between the states with $l=1$ and $l=0$
\cite{Put,Rokhsar,Leggett}. Recalling that when $l=1$, there is
a vortex state that is located at the center of the torus,
while for $l=0$ the vortex is at an infinite distance away, in
order for the vortex to exit the torus, there has to be a node
in the density of the atoms. However, this node costs
interaction energy, if the coupling is strong enough (since for
repulsive interactions the interaction energy is minimized for
a homogeneous density). The energy barrier that separates the
rotating state with $l=1$ from the lowest-energy state with
$l=0$ thus allows for the existence of persistent currents,
making the timescale for the decay of the current exponentially
long.

\begin{figure}[t]
\includegraphics[width=7cm,height=4.7cm]{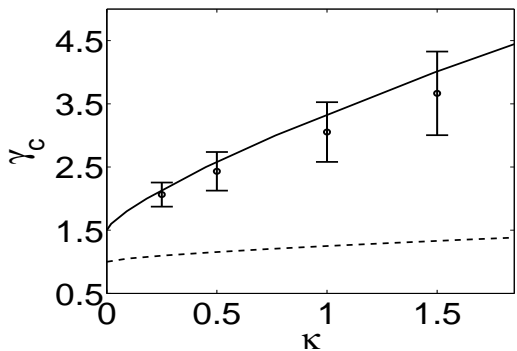}
\caption[]{The dots show the average value $\langle \gamma_c
\rangle$ and the bars show the standard deviation
$\sigma(\gamma_c)$, of the critical coupling $\gamma_c$, versus
the ``strength" of the random potential $\kappa$. These values
are calculated from the examination of 1000 different random
potentials. For each specific potential, the values of
$V(\theta)$ at each subinterval are chosen independently from a
uniform random distribution. The solid curve shows $\gamma_c$
for the specific potential shown in Fig.\,1. The dashed curve
shows $\gamma_c$ that results from Eq.\,(\ref{toy}), i.e., from
the toy model described in the text, for the same potential of
Fig.\,1.}
\label{FIG2}
\end{figure}

\section{Persistent currents within the mean-field
approximation}

We start our analysis of the stability of persistent currents
within the mean-field approximation, for relatively strong
disorder, where ${\tilde V}_0 \sim \hbar^2/(2 M R^2) \sim n_0 U_0$. In
the presence of any spatially-dependent potential $V(\theta)$,
the angular momentum is not a good quantum number, and even
more so in this case, where the disorder is not treated
perturbatively. Still, one may examine the energetic stability
of a state with one unit of circulation \cite{rem}; energetic
stability also guarantees dynamic stability \cite{JKL}. Clearly
such a state can only be metastable and not the lowest-energy
state of the system, since the non-rotating state has a lower
energy.

Before we describe our numerical results, it is instructive to
present a toy model that describes the problem qualitatively.
We thus consider the random potential shown in the lowest graph
of Fig.\,1, multiplied by some dimensionless constant $\kappa$,
i.e., $\kappa V(\theta)$; essentially $\kappa$ is the
``strength" of the disorder. Although this argument can be
generalized, for the sake of simplicity we consider a (very
limited) truncated order parameter
\begin{equation}
  \Psi_{\rm tr}(\theta) = \sqrt{1-l} \Phi_0 + e^{i \lambda} \sqrt l
\Phi_1,
\label{tr}
\end{equation}
that has an expectation value of the angular momentum per atom
equal to $l$. Here $\Phi_m = e^{i m \theta}/\sqrt{2 \pi}$ and
$\lambda$ is a phase that is determined below. This order
parameter gives an energy per atom that satisfies the equation
\begin{eqnarray}
\frac {\cal E} N - \frac {\gamma} 2 - \kappa V_0 = (1+\gamma)l
- \gamma l^2 - 2 \kappa |V_c| \sqrt{l (1-l)},
\label{toy}
\end{eqnarray}
where $V_c = \int_{-\pi}^{\pi} V(\theta) \cos \theta \, d
\theta/(2 \pi)$, and $V_0 = \int_{-\pi}^{\pi} V(\theta) \, d
\theta/(2 \pi)$ is the zero component of the Fourier transform
of $V(\theta)$. If $V_c$ is negative/positive, the phase
$\lambda$ is chosen to be $\lambda = 0 / \pi$, in order for the
last term in Eq.\,(\ref{toy}) to be always negative. Since,
according to Eq.\,(\ref{toy}), when $\kappa=0$, ${\cal E}(l)$
develops a local minimum at $l=1$ for a value of $\gamma = 1$,
the critical value of the coupling $\gamma_c$ that gives
metastability of the current is $\gamma_c = 1$ for the specific
$\Psi_{\rm tr}$. According to Eq.\,(\ref{toy}), for $\kappa
\neq 0$, the last term may destabilize the current, forcing
$\gamma_c$ to increase with $\kappa$, as shown in the dashed
curve of Fig.\,2. In other words, the coupling has to be
sufficiently strong in order for the state with nonzero
circulation to be stable. Clearly this curve is qualitatively,
but not quantitatively correct.

Turning to the numerical results, we attack this problem via
the method of imaginary time propagation \cite{relax}. We
choose different realizations of random potentials of the form
shown in Fig.\,1, multiplied by $\kappa$. For each specific
potential, the values of $V(\theta)$ at each subinterval are
chosen independently from a uniform random distribution. For
each specific $V(\theta)$, we identify the critical value of
the coupling $\gamma_c$ that is necessary to obtain stability
for a state with unit circulation for a specific $\kappa$,
starting with $\Phi_1$ as an initial condition. The solid curve
in Fig.\,2 shows $\gamma_c(\kappa)$ for the specific random
potential shown in Fig.\,1. For $\kappa \rightarrow 0^+$, our
results imply that $\gamma_c=3/2$; this result is analyzed
further below. We also perform a statistical analysis of the
values of $\gamma_c$ that we get using this method, i.e., we
compute the average value $\langle \gamma_c \rangle$ and the
standard deviation $\sigma(\gamma_c)$, for 1000 different
random potentials for each value of $\kappa$. Figure 2 shows
the result of these calculations. The values of $\gamma_c$ that
we get from the random potentials that we consider are
approximately Gaussian distributed.

The states that come out of our calculation are different than
$\Phi_1$ or $\Phi_0$, as they are not homogeneous, because of
the (random) potential. Figure 3 shows the calculated density
of the gas, for a circulation equal to zero (solid curve) and
unity (dashed curve), for the specific random potential shown
in Fig.\,1 (with $\kappa=1$). To get these densities, we use
the initial condition $\Phi_0 = 1/\sqrt{2 \pi}$ and
$\Phi_1=e^{i \theta}/\sqrt{2 \pi}$, respectively.

\begin{figure}[t]
\includegraphics[width=7.2cm,height=4.5cm]{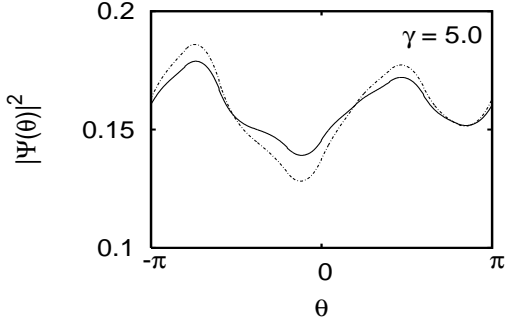}
\caption[]{The density $|\Psi(\theta)|^2$ of the cloud that
results from the imaginary-time propagation, for $\gamma=5$,
for the specific potential of Fig.\,1, with $\kappa=1$. The
solid/dashed curve corresponds to the state with zero units/one
unit of circulation.}
\label{FIG3}
\end{figure}

\section{Bogoliubov approach for weak disorder}

Another way to attack this problem, incorporating the
interactions, as well as the disorder, is via the Bogoliubov
transformation \cite{GMK}. This method gives the whole
excitation spectrum and also the depletion of the condensate,
which is assumed to reside at the single-particle state $\Phi_1
= e^{i \theta}/\sqrt{2 \pi}$. This assumption, as well as the
assumption for the existence of a Bose-Einstein condensate
requires that the disorder is not too strong.

Following the usual tricks, we replace the creation and
annihilation operators of atoms with angular momentum $m=1$,
$c_1^{\dagger}$ and $c_1$, with $\sqrt N_1$, where $N_1$ is the
number of atoms occupying the $m=1$ state. Then, since $N = N_1
+ \sum_{m \neq 1} c_m^{\dagger} c_m$, one can write for the
Hamiltonian
\begin{eqnarray}
   \hat{H} - N - \gamma N/2 = \sum_{m > 0} (m^2-2m+\gamma)
       c_{1-m}^{\dagger} c_{1-m}
\nonumber \\
 + (m^2+2m+\gamma) c_{1+m}^{\dagger} c_{1+m} + \gamma (c_{1-m}
c_{1+m} + c_{1-m}^{\dagger} c_{1+m}^{\dagger})
\nonumber \\
   + \epsilon_m^* (c_{1-m} + c_{1+m}^{\dagger})
   + \epsilon_m (c_{1-m}^{\dagger} + c_{1+m}),
\end{eqnarray}
where $\epsilon_m = \sqrt N V_m$. The last four terms describe
the processes where atoms scatter between the condensate and
the states $\Phi_{1\pm m}$ because of the (random) potential,
whose Fourier transform is denoted as $V_m$. We define the new
operators
\begin{eqnarray}
  \alpha_m &=& u_m c_{1-m} + v_m c_{1+m}^{\dagger}
  + \epsilon_m \frac {(m+2)u_m+(m-2)v_m}
  {m (m^2+2\gamma-4)}
 \nonumber \\
  \beta_m &=& u_m c_{1+m} + v_m c_{1-m}^{\dagger}
  + \epsilon_m^* \frac {(m-2)u_m+(m+2)v_m}
  {m (m^2+2\gamma-4)},
  \nonumber \\
\label{doef1}
\end{eqnarray}
and set $u_m = \cosh \theta_m$ and $v_m = \sinh \theta_m$, with
$\tanh(2\theta_m) = \gamma /(m^2 + \gamma)$, in order to
eliminate the off-diagonal terms. Then, the Hamiltonian is
written in the diagonal form
\begin{eqnarray}
 \hat H - N - \gamma N/2
  = \sum_{m > 0} (-2m + m \sqrt{m^2 + 2 \gamma}) \alpha_m^{\dagger}
 \alpha_m
 \nonumber \\
 + (2m + m \sqrt{m^2 + 2 \gamma}) \beta_m^{\dagger} \beta_m
\nonumber \\
 + m \sqrt{m^2 + 2 \gamma}
  - (m^2 + \gamma)
  - \frac {2 N |V_m|^2} {m^2+2 \gamma-4}.
\label{dham}
\end{eqnarray}

In order for the excitation spectrum to be positive, the
coefficient of the operator $\alpha_m^{\dagger} \alpha_m$
implies that $\gamma > 2 - m^2/2$. The most unstable mode is
the one with $m=1$, and therefore the critical coupling
constant for the existence of persistent currents is $\gamma_c
= 3/2$. This result was first derived in
Refs.\,\cite{GMK,Ueda}, within the truncated basis set of the
states with $m=0, 1$, and 2. According to the argument given
above, this is the exact value of $\gamma_c$ for the stability
of persistent currents, as we have also found numerically
within the mean-field approximation.

The eigenstates $|n_{\alpha, m}, n_{\beta, m} \rangle$ of the
diagonalized Hamiltonian of Eq.\,(\ref{dham}) (i.e., of the
number operators $\alpha^{\dagger}_m \alpha_m$ and
$\beta^{\dagger}_m \beta_m$) carry an angular momentum $L = N -
\sum_{m>0} m (n_{\alpha,m} - n_{\beta,m})$, i.e., $L \sim N \pm
{\cal O}(1)$. From Eq.\,(\ref{dham}) we also see that the
disorder potential enters the Hamiltonian only via the last
term, which can also be derived perturbatively. As a result,
one cannot see the dependence of $\gamma_c$ on $\kappa$ within
this method. In order to see this dependence, one would have to
consider a condensate of non-uniform density, or equivalently a
condensate with at least two single-particle states
macroscopically-occupied, e.g., of the form $|0^{N_0}, 1^{N_1}
\rangle$, as in Eq.\,(\ref{tr}).

Equation (\ref{dham}) can also give the depletion of the
condensate $\sum_{m > 0} \langle n_{\alpha,m} = 0, n_{\beta,m}
= 0 | c^{\dagger}_{1-m} c_{1-m} + c^{\dagger}_{1+m} c_{1+m} |
n_{\alpha,m} = 0, n_{\beta,m} = 0 \rangle$, which consists of
two parts, the one resulting from the interactions and the
other resulting from the disorder,
\begin{eqnarray}
 \Delta N_{\rm int} &=& \sum_{m > 0} 2 v_m^2 =
 \sum_{m > 0} \frac {\gamma + m^2} {m \sqrt{m^2 + 2 \gamma}} - 1,
\nonumber \\
  \Delta N_{\rm disorder} &=&
  2 N \sum_{m > 0} \frac {|V_m|^2 (m+2)^2} {m^2 (m^2+2\gamma-4)^2}.
\label{depint}
\end{eqnarray}

\section{Concluding remarks}

According to the results of the present study, one may conclude
very generally that disorder destabilizes states with nonzero
circulation. Physically, any kind of disorder results in an
inhomogeneous density distribution of the atoms. As compared to
the density $n_0$ of the atoms in the absence of any disorder
(which is homogeneous for repulsive interactions), the minimum
and maximum density, $n_{\rm min}$ and $n_{\rm max}$, in the
presence of disorder is thus both higher, as well as lower than
$n_0$, $n_{\rm min} < n_0 < n_{\rm max}$, as the conservation
of the number of atoms implies.

In other words, in the presence of disorder, there exist points
along the torus where the local density $n(\theta)$ is lower
than $n_0$. These points (where the density is lower than
$n_0$) give the cloud the chance to get rid of its circulation
at a lower energy expense as compared to the homogeneous case.
Higher values of $\kappa$ enhance the inhomogeneities in the
density, and thus make it easier for the vortex to slip out of
the torus. Therefore, the higher the strength of the disorder,
the higher the interaction strength that is necessary to make
the compressibility high enough, in order for the current to
become stable. The above arguments are rather general and apply
to any external potential.

The situation we consider here may be realized experimentally
in two different ways: (i) the gas is prepared in a rotating
state at a temperature above the condensation temperature, and
then is cooled down to zero temperature \cite{Corn}, or (ii)
starting from a gas at zero temperature, a phase is imprinted
\cite{Ketterle}. Finally, one could get evidence for the presence 
of current/circulation in such systems by, for example,
interference experiments \cite{Gupta}.

\acknowledgements

We thank A. D. Jackson and S. M. Reimann for useful
discussions, and J. \"Ogren for technical support. GMK
acknowledges financial support from the European Community
project ULTRA-1D (NMP4-CT-2003-505457).


\begin{thebibliography}{99}
\bibitem{rem} Although the gas may support states with higher
(integer) values of the circulation, we have examined the state
with one unit of circulation just as an example. In addition,
states of higher circulation would require a higher coupling
constant to become stable.
\end{thebibliography}
\end{document}